\documentstyle[12pt,amsfonts]{article}
\def\one{1\hskip-.37em 1}

\def\half{\textstyle{\frac{1}{2}}}
\def\quarter{\textstyle{\frac{1}{4}}}

\def\H{{\cal H}}
\def\ep{\epsilon}
\def\D{{\cal D}}

\def\E{{\rm I}\hskip-.2em{\rm E}}
\def\ra{\rightarrow}
\def\tint{{\textstyle\int}}

\def\s{\hskip.08em}
\def\d{\partial}
\def\l{\lambda}

\def\a{\alpha}
\def\b{\begin{eqnarray*}}  
\def\e{\end{eqnarray*}}    
\def\bn{\begin{eqnarray}}  
\def\<{\langle}
\def\>{\rangle}

\def\{{\lbrace}
\def\}{\rbrace}

\begin{document}
\title{Path Integrals, and \\Classical and Quantum Constraints\footnote{Based on a presentation at the 8th International Conference on Path Integrals, PI2005, Prague, Czech Republic, June, 2005.}}
\author{John R.~Klauder\\
Department of Physics and Department of Mathematics\\
University of Florida, Gainesville, FL 32611}
\date{}    
\maketitle
\begin{abstract}
Systems with constraints pose problems when they are quantized. Moreover, the Dirac procedure
of quantization prior to reduction is preferred. The projection operator method of
quantization, which can be most conveniently described by coherent state path integrals,
enables one to directly impose a regularized form of the quantum constraints. This procedure 
also overcomes conventional difficulties with normalization and second class constraints that
invalidate conventional Dirac constraint quantization procedures.
\end{abstract}
\section{Introduction}
In order to discuss the quantization of systems with constraints it is first important to briefly 
review what are two absolutely essential features of the very process of quantization itself.
First of all, we hold it self evident that: 

{\bf 1.} {\it The abstract operator formulation of quantum mechanics 
is correct and fundamental.} 

As a corollary of this viewpoint we next observe that:

{\bf 2.} {\it In order to properly describe quantum mechanics, it is necessary that any 
functional representation of quantum mechanics have an associated underlying operator
formulation.}

In particular this second property applies to:\vskip.1cm 
a) The Schr\"odinger partial differential
equation formulation of quantum mechanics,\vskip.1cm
and to: \vskip.1cm
b) Any version of a path integral formulation of quantum mechanics.\vskip.2cm

Although this paper is concerned with path integrals, it is pedagogically useful to spend
a few paragraphs on how these principles apply to the Schr\"odinger equation. The operator
form of this equation is given (in units where $\hbar=1$) by
  \b  i\s\d\Psi(t)/\d t = \H(t)\s\Psi(t)\;,  \e
where $\Psi(t)$ denotes the time dependent abstract vector $\Psi\in{\frak H}$, the abstract
Hilbert space, and $\H(t)$ denotes the (possibly) time-dependent, self-adjoint Hamiltonian operator.
As an example suppose that the system in question is a certain anharmonic oscillator characterized by
the fact that
   \b \H = \half(P^2+Q^2)+\lambda\s Q^4\;,  \e
where $P$ and $Q$ denote abstract, irreducible, self-adjoint Heisenberg operators that satisfy not only the 
Heisenberg commutation relation $[Q,P]=i\one$, but they also satisfy the
Weyl form of these relations, namely that
  \b e^{ipQ}\s e^{-iqP}=e^{ipq}\s e^{-iqP}\s e^{ipQ}\;, \e
for all real $c$-numbers $p$ and $q$. It was shown by von Neumann \cite{neu} that, apart from unitary equivalence,
there is only one realization of the operators $P$ and $Q$, namely, the Schr\"odinger representation
$P\ra -i\,\s\d/\d x$ and $Q\ra x$, acting on the Hilbert space $L^2({\bf R})$ of functions 
$\psi(x)$, $x\in{\bf R}$. Substitution of this representation for $P$ and $Q$ into the abstract
operator form for $\H$ yields the usual Schr\"odinger equation for this example, namely
  \b  i\s\d\psi(x,t)/\d t = -\half\psi^{''}(x,t)+\half x^2\psi(x,t)+\lambda\s x^4\psi(x,t)\;.  \e
So much for the obvious associations that apply to the Schr\"odinger equation.

However, it is useful to enquire what may happen if the connection to the operator formalism
is broken. The classical Hamiltonian for the anharmonic oscillator is normally taken to be
   \b H(p,q)=\half(p^2+q^2)+\lambda q^4\;,  \e
but after a canonical coordinate transformation of a suitable kind it is possible to express
the classical Hamiltonian for the same system in the form
   \b {\bar H}({\bar p},{\bar q})=\half{\bar p}^2\;,  \e 
or in still other coordinates in the form
   \b  {\tilde H}({\tilde p},{\tilde q}) = \half({\tilde p}^2+{\tilde q}^2)\;,  \e
etc. All of these functionally unequal forms properly describe the same physical system in the indicated canonical coordinates. Clearly,
to promote the coordinates in these distinct cases to canonical Heisenberg operators would lead
to Hamiltonian operators with quite different spectra and they all can not be physically correct. How
is one to know which set of canonical coordinates to promote to canonical operators so as to
obtain the correct physical spectrum for the specific system under consideration? The answer,
according to Heisenberg, Schr\"odinger, and Dirac \cite{dir}, is that the classical canonical
coordinates should be chosen as ``Cartesian coordinates". 

To put some further substance in this remark, it is useful to appeal to coherent states \cite{klbo}. 
In particular, let $|0\>$ denote a normalized vector that satisfies $(Q+iP)\s|0\>=0$, namely,
$|0\>$ is the ground state of an harmonic oscillator with unit angular frequency. Let
  \b |p,q\>\equiv e^{-iqP}\s e^{ipQ}\s|0\> \;, \e
for all $(p,q)\in {\bf R}^2$, denote the set of coherent states. Then, in view of the Heisenberg
commutation relation, it follows for a general Hamiltonian operator $\H(P,Q)$ that
   \b && H(p,q)\equiv\<p,q|\s\H(P,Q)\s|p,q\>\\
   &&\hskip1.43cm=\<0|\s\H(P+p,Q+q)\s|0\>\\
&&\hskip1.43cm=\H(p,q)+{\cal O}(\hbar;p,q)\;; \e
the last form of this expression is particularly evident for polynomial Hamiltonians. In any
case, in the chosen coordinates for the Weyl group, apart from explicitly $\hbar$ dependent
terms, the $c$-number Hamiltonian, $H(p,q)$, defined above, has the same functional form as the $q$-number Hamiltonian, $\H(p,q)$. In other group coordinates that would not be the case, generally speaking. Thus
to associate a particular expression for a classical Hamiltonian to the proper quantum Hamiltonian operator, one needs to use Cartesian coordinates. How can we call our choice
of coordinates ``Cartesian"? This association follows from the Fubini-Study metric induced on phase space by the coherent states, namely, by
the fact that
  \b 2[\s|\!|\s d|p,q\>\s|\!|^2-|\<p,q|\s d|p,q\>|^2\s] = dp^2+dq^2 \e
in the indicated choice of group coordinates.
 
\section*{Path integrals for systems without constraints}
We now take up the question of path integrals and for pedagogical purposes we start with the
simpler and more familiar situation in which there are no constraints. The abstract operator
solution to Schr\"odinger's equation for a time-dependent Hamiltonian may be written as
   \b \Psi(T) ={\sf T}\s e^{-i\s\tint_0^T\s\H(t)\s dt}\,\Psi(0)\;,  \e
where ${\sf T}$ denotes time ordering. For sufficiently smooth time dependence, the evolution operator
  \b  U(T)={\sf T}\s e^{-i\s\tint_0^T\s\H(t)\s dt} \e
may be represented as the limit of a large number of small time steps, namely, as
  \b U(T)=\lim_{N\ra\infty}\s e^{-i\ep\H_N}\s\cdots\s e^{-i\ep\H_2}\s e^{-i\ep\H_1}\;, \e
where $\ep\equiv T/N$, $T>0$,  and $\H_k\equiv \H(k\s\ep)$. This formula may be put to good use in at
least two different ways.
\subsection*{Phase space path integral -- case A}
First, to form the propagator between (formal) sharp position states $|q\>$, where $Q\s|q\>=q\s|q\>$, for all $q\in{\bf R}$, let us insert repeated resolutions 
of unity as customary to yield
  \b \<q{''}|\s U(T)\s|q'\>=\lim_{N\ra\infty}\int\cdots\int\,\Pi_{n=0}^N\,\<q_{n+1}|\s e^{-i\ep\H_n}\s|q_n\>\,\Pi_{n=1}^N\s dq_n\;,  \e
where $q''=q_{N+1}$ and $q'=q_0$. As a next step we can insert resolutions of unity over the
conjugate momentum states to give
  \b &&\<q{''}|\s U(T)\s|q'\>=\lim_{N\ra\infty}\int\cdots\int\,\Pi_{n=0}^N\,\<q_{n+1}|p_{n+1/2}\>\<p_{n+1/2}|\s e^{-i\ep\H_n}\s|q_n\>\\
&&\hskip4.96cm\times\Pi_{n=0}^N\s dp_{n+1/2}\,\Pi_{n=1}^N\s dq_n \;.\e
To emphasize that we are using two different resolutions of unity, which requires diagonalizing both operators $Q$ and $P$, and which can only be done at different times, i.e., sequentially, we have used the notation $|q_n\>$ and $|p_{n+1/2}\>$.

If we introduce the fact that
  \b \<q_{n+1}|p_{n+1/2}\>=\frac{e^{i\s p_{n+1/2}\s q_{n+1}}}{\sqrt{2\pi}}\;,  \e
as well as expand each exponential to first order in $\ep$, we are led to the familiar expression for the sharp $q$ to sharp $q$ propagator for the phase space path integral given by  
  \b &&\hskip-.5cm{\cal M}\int\,e^{i\tint[p\s{\dot q}-H(p,q)]\s dt}\,\D p\,\D q \\
 &&\hskip0cm =\lim_{N\ra\infty}\int\cdots\int\,\Pi_{n=0}^N\,e^{i\s p_{n+1/2}\s(q_{n+1}-q_n)}\,[1-i\s\ep\s\<p_{n+1/2}|\s\H_n\s|q_n\>/\<p_{n+1/2}|q_n\>]\\
&&\hskip2.6cm\times\Pi_{n=0}^N\s dp_{n+1/2}/(2\pi)\,\Pi_{n=1}^N\s dq_n \;.\e
It is in this familiar way that meaning can be given to the formal phase space path integral
through a close association with the abstract operator formulation. Of course, implicit in the expression for the overlap $\<q_{n+1}|p_{n+1/2}\>$ is the assumption of Cartesian coordinates. 

While this expression is mathematically correct for a wide class of Hamiltonians, it is nevertheless important to point out that it is ``unnatural" from a physical point of view since it asserts that the phase space ``paths" involved repeatedly oscillate between sharp $q$ (and thereby absolutely {\it no} knowledge of $p$) and sharp $p$ (and thereby absolutely {\it no} knowledge of $q$). 
\subsubsection*{Phase space path integral -- case B}
We can derive another expression for the meaning of the ``same" phase space path integral in the following way. Rather than alternately use sharp $p$ and sharp $q$ states, let us repeatedly use just coherent states and their associated resolution of unity. As a consequence, the same initial expression
  \b U(T)=\lim_{N\ra\infty}\s e^{-i\ep\H_N}\s\cdots\s e^{-i\ep\H_2}\s e^{-i\ep\H_1} \e
leads to
  \b &&\hskip-.3cm\<p'',q''|\s U(T)\s|p',q'\>\\
&&\hskip.3cm=\lim_{N\ra\infty}\int\cdots\int\,\Pi_{n=0}^N\,\<p_{n+1},q_{n+1}|\s e^{-i\ep\H_n}\s|p_n,q_n\>\,\Pi_{n=1}^N\,dp_ndq_n/(2\pi)\;, \e
where $p''=p_{N+1},\,q''=q_{N+1}$ and $p'=p_0,\,q'=q_0$. If we use the fact that
 \b &&\hskip-.3cm\<p_{n+1},q_{n+1}|p_n,q_n\>\\
&&\hskip.3cm=\exp\{\s i\half(p_{n+1}+p_n)(q_{n+1}-q_n)-\quarter[(p_{n+1}-p_n)^2+(q_{n+1}-q_n)^2]\s\}\;, \e
as well as expand the exponential to first order in $\ep$, as before, we are led to an 
alternative, coherent state representation, for the formal phase space path integral given by 
  \b &&\hskip-.5cm{\cal M}\int\,e^{i\tint[p\s{\dot q}-H(p,q)]\s dt}\,\D p\,\D q \\
  &&\hskip0cm = \lim_{N\ra\infty}\int\cdots\int\Pi_{n=0}^N\,e^{\{\s i(p_{n+1}+p_n)(q_{n+1}-q_n)/2-[(p_{n+1}-p_n)^2+(q_{n+1}-q_n)^2]/4\s\}}\\
&&\hskip.5cm\times[1-i\s\ep\s\<p_{n+1},q_{n+1}|\s\H\s|p_n,q_n\>/\<p_{n+1},q_{n+1}|p_n,q_n\>]
  \;\Pi_{n=1}^N\,dp_n\,dq_n/(2\pi)\;.  \e
One again, this expression is based on the implicit use of Cartesian coordinates.

Unlike case A above, this version of the phase space path integral is both mathematically
correct for a large class of Hamiltonians as well as being physically ``natural". It is natural because
the meaning of the variables $p$ and $q$ is that of {\it mean values} rather than sharp values, and it is perfectly legitimate to specify the mean values of both $p$ and $q$ at equal times -- and do so for all time. The 
meaning of these variables as mean values stems from the fact that $\<p,q|\s P\s|p,q\>=p$ and
$\<p,q|\s Q\s|p,q\>=q$.

{\bf Remark:} Although case A and case B led to quite different results starting from the same formal expression, it is noteworthy that they both made use of a first-order expansion of the exponential in the parameter $\ep$. In particular, in both cases we made use of the approximation
  \b  e^{-i\ep\H_k}\simeq 1-i\ep\s\H_k \;.  \e
When it comes to deal with constraints, it will become clear that this approximation for the constraints is insufficient.

\section*{Classical theory of constraints -- a sketch}
In order to account for constraints, it is only necessary to augment the usual classical
action functional by the addition of the constraints along with Lagrange multipliers. The
result is an action functional given generically by the expression
   \b I=\tint[\s p_j\s{\dot q}^j-H(p,q)-\l^\a\s\phi_\a(p,q)\s]\,dt\;. \e
Here, $1\le j\le J$ and $1\le\a\le A$, where $J$ and $A$ denote the numbers of canonical degrees of freedom $(p_j,q^j)$ and constraints $\phi_\a(p,q)$, respectively, while $\l^\a(t)$ denotes the several Lagrange
multipliers. Variation of $p_j$, $q^j$, and $\l^\a$ lead to the basic equations, namely,
  \b && \hskip1.0cm{\dot q}^j=\d H(p,q)/\d p_j+\l^\a\s\d\phi_\a(p,q)/\d p_j\;,  \\
   && \hskip1.0cm{\dot p}_j=-\d H(p,q)/\d q^j-\l^a\s\d\phi_\a(p,q)/\d q^j\;, \\
    && \phi_\a(p,q)=0\;.  \e
The subset of phase space on which the constraints holds is called the constraint hypersurface.
The equations of motion may also be written in terms of Poisson brackets. In particular, since the constraints must hold for all time, it is necessary that
  \b {\dot\phi}_\a(p,q)=0=\{\phi_\a(p,q),H(p,q)\}+\l^\beta\s\{\phi_\a(p,q),\phi_\beta(p,q)\} \e
holds on the constraint hypersurface.
This latter equation divides constraints into two principal classes. 

Suppose first that the Poisson brackets among the constraints vanish on the constraint hypersurface. In that case the second term is already zero for any choice of the Lagrange multipliers; the first term therefore also needs to vanish on the constraint hypersurface (or otherwise
it determines a new constraint that must be included). These conditions may be stated as
  \b &&\{\phi_\a(p,q),\phi_\beta(p,q)\}=c_{\a\beta}^{\;\;\;\;\gamma}\,\phi_\gamma(p,q)\;,\\
   &&\hskip.1cm\{\phi_\a(p,q),H(p,q)\}=h_\a^{\;\;\beta}\,\phi_\beta(p,q)\;.  \e
Constraints that fulfill such equations are called {\it first class constraints}. A further
division is made as follows: If the coefficients $c_{\a\beta}^{\;\;\;\;\gamma}$ are 
constants, the constraints are called closed first class constraints; if instead the coefficients $c_{\a\beta}^{\;\;\;\;\gamma}$ are general functions of phase space, then the constraints are called open first class constraints. Moreover, to solve the equations of motion it is generally necessary that some specific choice of the Lagrange multipliers be made; this is called
a choice of gauge. Yang-Mills theories have closed first class constraints, while gravity is an
open first class system.

Next, let us suppose that the Poisson brackets of the constraints do {\it not} vanish on the
constraint hypersurface. For simplicity, let us even assume the case where the Poisson brackets of the constraints $\{\phi_\a(p,q),\phi_\beta(p,q)\}$ form an invertible matrix. In that case, the
Lagrange multipliers are fully determined and are given by
 \b\l^\beta\equiv\s-\s[\, \{\phi_\a(p,q),\phi_\beta(p,q)\}\,]^{-1}\;\{\phi_\a(p,q),H(p,q)\}\;. \e
Constraints that have such properties are called 
{\it second class constraints}.

Of course, there also exist mixed situations in which some of the constraints are first class while the rest are second class.
\section*{Constraint quantization -- reduction \\before quantization}
In this section we outline the well known procedures of Faddeev \cite{fad} and Senjanovi\'c \cite{sec} for dealing with first and second class constraint situations, respectively. We proceed
formally as is customary in such cases. Consider the formal phase space path integral
\b&&{\cal M}\int e^{i\tint[p_j\s{\dot q}^j-H(p,q)-\l^\a\s\phi_\a(p,q)]\s dt}\,\D p\,\D q\,\D\l\\
 &&\hskip1cm={\cal M}\int e^{i\tint[p_j\s{\dot q}^j-H(p,q)]\s dt}\delta\{\phi(p,q)\}\,\D p\,\D q \e
(modulo a redefinition of ${\cal M}$),
which enforces the classical constraints exactly. The resultant integral may well diverge (e.g.,
if $\phi_1=p_1$ and $H(p,q)$ is independent of $q^1$). Gauge fixing is used to overcome possible
divergences, and the Faddeev-Popov determinant is introduced to maintain formal covariance under canonical coordinate transformations. The path integral expression now reads
  \b {\cal M}\int e^{i\tint[p_j\s{\dot q}^j-H(p,q)]\s dt}\delta\{\chi(p,q)\}\s\det\{\chi^\a,\phi_\beta\}\delta\{\phi(p,q)\}\,\D p\,\D q\;, \e
where $\chi^\a(p,q)=0$, for all $\a$,  determines the gauge choice. This expression
is expected to be equal to 
  \b{\cal M}^*\int e^{i\tint[p^*_B\s{\dot q}^{*B}-H^*(p^*,q^*)]\s dt}\,\D p^*\,\D q^*  \e
where $B$ is an index that runs over the remaining, ``physical" degrees of freedom, $p^*$ and $q^*$. The formulation given above formally applies to the case of first class constraints.

In a case of purely second class constraints, the final result is taken to be
 \b{\cal M}\int e^{i\tint[p_j\s{\dot q}^j-H(p,q)]\s dt}\,[\s\det\{\phi_\a,\phi_\beta\}\s]^{1/2}\,\delta\{\phi(p,q)\}\,\D p\,\D q \;,  \e
which again is formally equivalent to an expression of the sort
\b{\cal M}^*\int e^{i\tint[p^*_B\s{\dot q}^{*B}-H^*(p^*,q^*)]\s dt}\,\D p^*\,\D q^* \;. \e

The foregoing expressions are plausible, formal phase space path integrals, but -- and, in the author's opinion, this is an important qualification -- these path integral expressions have lost any direct connection with an underlying abstract operator approach. While they surely can be used to calculate results, and on many occasions the results may well be correct, there simply is no firm foundation tied to an abstract operator approach to ensure that the results will be universally valid.

To rectify that situation we first need to remind ourselves what is the accepted abstract operator formulation of quantization when constraints are present.
\section*{Abstract operator quantization with \\constraints -- quantization before reduction}
The general abstract operator quantization procedure for systems with constraints is due to
Dirac \cite{dir2}. In this approach one quantizes first and reduces second. This is the preferred order since one then has the chance to employ Cartesian coordinates in the quantization, which,
as described earlier, is the proper set of coordinates to promote to canonical operators. (Reduction first may give rise to a constraint hypersurface that does not admit Cartesian coordinates.) Thus,
we suppose that we have obtained suitable canonical operators $Q^j$ and $P_j$, $1\le j\le J$,
and also chosen an acceptable factor ordering, if necessary, such that the Hamiltonian $\H(P,Q)$ and the several constraint operators $\Phi_\a(P,Q)$ are self adjoint operators. Reduction
consists in seeking a Hilbert space, ${\frak H}_{phys}$, called the physical Hilbert space, which
is a subspace of the original Hilbert space ${\frak H}$, i.e., ${\frak H}_{phys}\subset{\frak H}$. The
elements of ${\frak H}_{phys}$ are those Hilbert space vectors for which
 \b \Phi_\a(P,Q)\s\Psi_{phys}=0  \e
for all $\a$, $1\le\a\le A$. Clearly, such vectors form a linear space. However, there are
two special issues that must be considered. First, it follows from this criterion that
  \b [\Phi_\a(P,Q),\s\Phi_\beta(P,Q)]\s\Psi_{phys}=0\;, \e
but this condition may have $\Psi_{phys}=0$ as its only nontrivial solution. This situation arises
for second class constraint systems. To deal with that, Dirac restricts his procedure to suitable first class systems; second class systems are dealt with in a completely different manner. Second, it
may happen that the only nontrivial solutions are formal eigenvectors in the sense that $(\Psi_{phys},\s\Psi_{phys})=\infty$. If this is the case, then some procedure must be introduced to deal
with the fact that no true vectors exist that belong to ${\frak H}_{phys}$. This procedure is
not quite as straightforward as one might imagine.

In the next section we outline a relatively new procedure \cite{klau} to deal with quantum constraints
that is able to handle second class constraints as easily and with the same formalism as
first class constraints, as well as having a well defined procedure to deal with those cases that have formal eigenvectors that are not in Hilbert space. 

\section*{Projection operator method for \\quantum constraints}
Ideally, if $\Phi_\a\s\Psi_{phys}=0$ for all $\a$, it should follow that
  \b \Sigma_\a\s\Phi_\a^2\s\Psi_{phys}=0 \e
holds as well. This relation works sometimes but not always. Therefore, let us relax this
latter condition and replace it as follows. Assume that the operator $\Sigma_\a\s\Phi_\a^2$ is self adjoint
and has a spectral representation given by
   \b \Sigma_\a\s\Phi_\a^2=\int_0^\infty \l\,d\s\E(\l) \e
expressed in terms of the associated spectral family of projection operators $\{\s\E(\l):\s 0\le\l<\infty\s\}$.
We introduce the projection operator
  \b  \E(\Sigma_\a\s\Phi_\a^2\le\delta(\hbar)^2)\equiv \int_0^{\delta(\hbar)^2}\,d\s\E(\l)\;, \e
which projects onto the spectral interval from $0$ to $\delta(\hbar)^2$. Here, $\delta(\hbar)$
denotes a small parameter to be chosen appropriately; it is {\it not} a Dirac delta function!
Finally, the physical Hilbert space is given by 
  \b {\frak H}_{phys}\equiv \E\s{\frak H} \;.  \e
A few examples will help explain how the projection operator method works. 

First, let
$\Phi_k=J_k$, $k=1,2,3$, be the generators of the rotation group. We want to project onto
those states for which $J_k\s\Psi_{phys}=0$ for all $k$. We do so by considering
  \b \E =\E(J_1^2+J_2^2+J_3^2\le \hbar^2/2)\;. \e
Since $\Sigma_k J_k^2$ is just the Casimir operator for the rotation group, with eigenvalues
given by $j(j+1)\hbar^2$, $j=0,\half,1,\dots$, it follows that $j=0$ is the only subspace
allowed by the projection operator. (Clearly, a small range of other values for $\delta(\hbar)^2$
works just as well, but we shall not dwell on that aspect.)

Second, let $\Phi_1=P$ and $\Phi_2=Q$. The equations $P\s\Psi_{phys}=0$ and $Q\s\Psi_{phys}=0$
imply that $[Q,P]\s\Psi_{phys}=i\hbar\s\Psi_{phys}=0$, i.e., $\Psi_{phys}=0$. This is the classic example of a second class system for which the original Dirac procedure does not work. However,
let us choose
  \b \E=\E(P^2+Q^2\le \hbar) \e
which acts to project onto vectors for which $(Q+iP)\s\Psi_{phys}=0$. If $Q$ and $P$ are
irreducible, then the only solution is a projection onto the ground state of an harmonic oscillator with unit angular frequency. The essential point is the projection in this case is onto a one dimensional subspace.

It is noteworthy that the first example consists of an operator with a discrete spectrum that contains zero (first class system), while the second example involves an operator with a discrete spectrum that does {\it not} include zero (second class system).

Third, let $\Phi_1=P$ be the only constraint. This operator has its zero in the continuous
spectrum, and thus all nontrivial solutions to the equation $P\s\Psi_{phys}=0$ obey $(\Psi_{phys},\s\Psi_{phys})=\infty$. In the projection operator language, the operator
  \b \E=\E(P^2\le \delta^2)  \e
vanishes as $\delta\ra0$, so care must be taken to extract the ``germ" of this limit. (An $\hbar$ dependence is not important in this case.) To extract the
desired ``subspace" where ``$P=0$", it is most convenient to adopt a representation space.
For that purpose let us choose a coherent state basis. In particular, let us consider the quotient
\b &&\<p'',q''|\E(P^2\le\delta^2)|p',q'\>\s{\bigg/}\s\<0|\E(P^2\le\delta^2)|0\>\\
 && \hskip.3cm=\int_\delta^\delta e^{-(k-p'')^2/2+ik(q''-q')-(k-p')^2/2}\,dk\s{\bigg/}\s\int_\delta^\delta e^{-k^2}\,dk\;.  \e
As $\delta\ra0$, the numerator and the denominator each vanish; however, the quotient will not vanish. Indeed, as $\delta\ra0$, this quotient becomes
 \b  e^{-(p''^2+p'^2)/2}\;,  \e
which characterizes a {\it one} dimensional physical Hilbert space, which is a perfectly acceptable result in this case. Since this expression no longer depends on $q''$ or $q'$, it is clear that we have reached the space where ``$P=0$". Observe that the physical Hilbert space in this case is, strictly speaking, not a subspace of the original Hilbert space ${\frak H}$.
Nevertheless, from a representation point of view, it is important to observe that the physical Hilbert space of interest can be obtained by a suitable limit taken from within the original Hilbert space ${\frak H}$.
\subsubsection*{Dynamics}
There are two important cases when dynamics is considered. The first case assumes that the Hamiltonian is an {\it observable}. An observable operator ${\cal O}$ is one which commutes with the projection operator; specifically, that
   \b [{\cal O},\s\E]=0 \;. \e
Therefore, if the Hamiltonian is an observable, it follows that
   \b  [\H,\s\E]=0 \;. \e
In that case, we clearly have the operator identity that
  \b  e^{-i\H T}\s\E=\E\s e^{-i(\E\H\E)T}\s\E \;.  \e
This equation asserts that when $\H$ is an observable and commutes with the projection operator, it is sufficient to impose the projection operator at just one time -- here chosen as the initial time -- and then the temporal evolution remains thereafter within the physical subspace, and, moreover,  the temporal evolution is generated by that component of the Hamiltonian that lies within the physical subspace. The Hamiltonian is an observable for first class systems and for those second class systems for which the Hamiltonian vanishes.

The second and more general situation is when the Hamiltonian is not an observable, namely, in cases for which
   \b [\H,\s\E]\neq0\;. \e
We would still like to ensure that the temporal evolution lies wholly within the physical subspace, and it is clear that one initial application of the projection operator will {\it not} be sufficient. Just as we use the classical Lagrange multipliers to force the time evolving classical system
back to the constraint hypersurface when we need to, we can use the projection operator to
force the time evolving quantum system back to the physical subspace when we need to. In symbols, this argument suggests that we consider
  \b \lim_{N\ra\infty}\s e^{-i\ep\H}\s\E\s\cdots e^{-i\ep\H}\s\E\s e^{-i\ep\H}\s\E \;, \e
where, as before, $\ep=T/N$, and $T>0$ is fixed. As shown by Chernoff \cite{him}, this limit
is exactly 
  \b \E\s e^{-i(\E\H\E)T}\s\E  \;,  \e
as desired. In the second class case, there are special examples where the temporal evolution is not unitary, e.g., if $\H=P$ and $\E$ is a projection onto the positive half line, $Q>0$. However, if the original Hamiltonian is bounded below, which is more common in physical situations, then there is always a unitary version of the desired temporal evolution in the physical subspace. 

It is clear that a first class system can also be treated with repeated alternate projections and short time propagations, so the procedure outlined for second class systems works equally well for all systems.

\subsubsection*{Integral representation for projection operator}
In special cases, such as first class systems that correspond to compact groups, it is
straightforward to find integral representations that yield an appropriate projection operator. However, it it noteworthy that there exists a universal integral representation that yields the desired projection operator for {\it any} set of constraint operators \cite{int}. We have in mind the {\it operator identity} given by 
  \b \E(\Sigma_\a\s\Phi_\a^2\le\delta(\hbar)^2)=\int {\sf T}\s e^{-i\tint_0^\tau\s\l^\a(t)\s\Phi_\a\,dt}\,\D R(\l)\;,  \e
which involves a time ordered functional integral over $c$-number 
Lagrange multipliers, where $R(\l)$ is
a suitable (weak) measure. This result holds for any $\tau>0$ (note that the left side is independent of $\tau$). The measure $R(\l)$ depends on $\tau$, $\delta(\hbar)^2$, and the number of constraints, but it is {\it totally independent} of the choice of the set of constraint operators $\{\Phi_\a\}$. Indeed, this expression applies even if the constraint operators all {\it vanish}, in which case we learn that
  \b 1=\int \D R(\l) \;.  \e

Such an integral representation for the projection operator can be explicitly used in forming a path integral representation for a system with constraints, and since the measure is
the same for all systems, it may be used to provide a common formulation for any constrained system. Since we use the explicit measure in the following section, we will not describe it here.

We now turn our attention to providing a phase space path integral formulation of temporal evolution in the presence of general constraints that maintains a close association with the abstract operator formulation that we have presented. 
\section*{Coherent state path integrals with constraints} 
We wish to find an interpretation of the formal phase space path integral
  \b {\cal M}\int e^{i\tint_0^T[\s p_j\s{\dot q}^j-H(p,q)-\l^a\phi_\a(p,q)]\s dt}\,\D p\,\D q\,\D R(\l) \e
that yields the desired expression
 \b \<p'',q''|\s\E\s e^{-i(\E\H\E)T}\s\E\s|p',q'\>  \e
for temporal propagation in the physical Hilbert space.

In the following equation chain, the weak measure $R(\l)$ is made explicit as we choose a formula that achieves our goal, namely:

\b &&{\cal M}\int e^{i\tint_0^T[\s p_j\s{\dot q}^j-H(p,q)-\l^a\phi_\a(p,q)]\s dt}\,\D p\,\D q\,\D R(\l) 
  \equiv\lim_{N\ra\infty}\int\cdots\int\prod_{n=0}^N\Bigg\{\\
&&\hskip.4cm\times\lim_{M\ra\infty}\,\int\cdots\int\prod_{m=1}^M\Bigg[\,\Bigg(\s\<p_{n+m/M},q_{n+m/M}|p_{n+(m-1)/M},q_{n+(m-1)/M}\>\\
&&\hskip.4cm+\delta_{m,M}\<p_{n+m/M},q_{n+m/M}|\s(-i\s\ep\s\H)\s|p_{n+(m-1)/M},q_{n+(m-1)/M}\>\\
&&\hskip.4cm+\<p_{n+m/M},q_{n+m/M}|\s[\s-i(\ep/M)\l^\a_{n,m}\s\Phi_\a-(\ep^2/2M^2)\l^\a_{n,m}\s\l^\beta_{n,m}\Phi_\a\s\Phi_\beta\s]\s\\
&&\hskip.4cm\times\s|p_{n+(m-1)/M},q_{n+(m-1)/M}\>\,\Bigg)\,(c\gamma_n)^{-A/2}\,e^{-i\ep\s/(4M\gamma_n)\s\Sigma_\a\s\l^{\a\s2}_{n,m}}\,\Pi_\a\,d\l^\a_{n,m}\Bigg]\,\Bigg\}\\
&&\hskip.4cm\times \prod_{n=1}^N\Bigg[\,\Bigg(\prod_{m=1}^M\;dp_{n+m/M}\s dq_{n+m/M}\,\Bigg)\,d\sigma(\gamma_n)\,\Bigg]\\
&&=\lim_{N\ra\infty}\int\cdots\int\prod_{n=0}^N\Bigg\{\<p_{n+1},q_{n+1}|(1-i\ep\H)\s e^{i\gamma_n\ep\s\Sigma_\a\s\Phi_\a^2}\s|p_n,q_n\>\\
&&\hskip.4cm\times\frac{\sin[\gamma_n\ep\delta(\hbar)^2]}{\pi\s\gamma_n}\,d\gamma_n\Bigg\}\prod_{n=1}^N\,dp_n\s dq_n/(2\pi)\\
&&= \lim_{N\ra\infty}\int\cdots\int\prod_{n=0}^N\<p_{n+1},q_{n+1}|\s e^{-i\ep\H}\s\E\s|p_n,q_n\>\;\prod_{n=1}^N\,dp_n\s dq_n/(2\pi)\\
&&=\lim_{N\ra\infty}\<p'',q''|\s e^{-i\ep\H}\s\E\s\cdots e^{-i\ep\H}\s\E\s e^{-i\ep\H}\s\E\s
|p',q'\>\\
&&=\<p'',q''|\s\E\s e^{-i(\E\H\E)T}\s\E\s|p',q'\>\;. \e
Here, as usual, $p'',q''=p_{N+1},q_{N+1}$ as well as $p',q'=p_0,q_0$. The constant $c=-4\pi iM/\ep$  part way
through the equation chain is a normalization chosen to ensure the form of the equation which follows the one in which $c$ appears.

It is important to observe that, unlike the Hamiltonian $\H$, it was necessary to expand the expression involving the constraints $\Phi_\a$ to {\it second} order in the small parameter
$\ep$. In addition, it was necessary to introduce an additional refinement ($M$) of each small time step ($\ep$) in order to construct a projection operator $\E$ to go along with each of the large number ($N$) of small time step evolutions for the Hamiltonian.

\section*{Summary}
With this final expression we have achieved our goal of providing a path integral formulation for canonical systems with general constraints that is closely associated with the abstract operator formulation. It is noteworthy that this formulation offers a path integral approach to the quantization of systems with first and second class constraints that does {\it NOT} involve: gauge fixing, Faddeev-Popov determinants, Gribov ambiguities, moduli space, auxiliary variables, ghosts, indefinite metrics, Dirac brackets, etc.


\begin{thebibliography}{99}
\bibitem{neu} J. von Neumann, Math. Ann. {\bf 104}, 570 (1931).

\bibitem{dir} See, e.g., P.A.M.~Dirac, {\it The Principles of Quantum Mechanics} (Clarendon Press, Oxford, 4th Edition, 1958), p.~114.

\bibitem{klbo} See, e.g., J.R.~Klauder and B.-S.~Skagerstam, {\it Coherent States}, (World Scientific, Singapore, 1985).

\bibitem{fad} L.D.~Faddeev, Theoret. Math. Phys. {\bf 1}, 1 (1970).

\bibitem{sec} P.~Senjanovi\'c, Ann. Phys. {\bf 100}, 227 (1976).

\bibitem{dir2} P.A.M.~Dirac, {\it Lectures on Quantum Mechanics} (Belfer Graduate School of Science, Yeshiva University, New York, 1964).

\bibitem{klau} J.R.~Klauder, Ann.~Phys.~ {\bf 254}, 419 (1997); ``Coherent State Path Integrals for Systems with Constraints'', in {\it Path Integrals: Dubna `96}, Proceedings of the International Seminar ``Path Integrals: Theory \& Applications'' and 5th International Conference on path Integrals from meV to MeV, Dubna, Russia, (Publishing Department, JINR, Dubna, Russia, 1996), p.~51. See also, S.V.~Shabanov, in {\it Path integrals: Dubna `96}, Proceedings of the International Seminar ``Path Integrals: Theory \& Applications'' and 5th International Conference on path Integrals from meV to MeV, Dubna, Russia, (Publishing Department, JINR, Dubna, Russia), p.~133.

\bibitem{him} P. Chernoff, J. Funct. Anal. {\bf 2}, 238 (1968).

\bibitem{int} J.R.~Klauder, Nucl.~Phys.~B {\bf 254}, 419 (1997).

\end{thebibliography}
\end{document}